# The...Tinderverse?: Opportunities and Challenges for User Safety in Extended Reality (XR) Dating Apps

Sarath S. Shanker
Computer Science and Engineering
Oakland University
Rochester, Michigan, USA
sshanker@oakland.edu

Douglas Zytko
Computer Science and Engineering
Oakland University
Rochester, Michigan, USA
zytko@oakland.edu

## ABSTRACT

Dating apps such as Tinder have announced plans for a dating metaverse: the incorporation of XR technologies into the online dating process to augment interactions between potential sexual partners across virtual and physical worlds. While the dating metaverse is still in conceptual stages we can forecast significant harms that it may expose daters to given prior research into the frequency and severity of sexual harms facilitated by dating apps as well as harms within social VR environments. In this workshop paper we envision how XR could enrich virtual-to-physical interaction between potential sexual partners and outline harms that it will likely perpetuate as well. We then introduce our ongoing research to preempt such harms: a participatory design study with sexual violence experts and demographics at disproportionate risk of sexual violence to produce mitigative solutions to sexual violence perpetuated by XR-enabled dating apps.

## 1 Introduction

In this workshop paper we highlight the emergence of the dating metaverse [7,8], or incorporation of XR technologies into tinder and other dating apps, such as social VR environments for potential sexual partners to interact before meeting face-to-face [2]. While XR provides ample potential for enriching interaction between potential romantic and sexual partners, it can also exacerbate sexual harms that dating apps are already known to perpetuate—and introduce new ones.

In the following sections we first introduce the concept of the dating metaverse and recent design progress from leading dating app companies such as Tinder. We then elucidate opportunities for XR to protect users during online-to-offline interaction with strangers as well as looming harms. We conclude by introducing our ongoing research into designing safety-conscious XR technologies for online dating, which we hope will spark discussion amongst workshop attendees regarding the role of XR in perpetuating and also mitigating interpersonal harm across virtual and physical world settings.

## 2 The Dating Metaverse

The notion of applying XR technologies to online dating to enrich interaction between potential romantic and sexual partners has been considered for some time. Previous works have evinced keen user interest and demand for platforms that convene intimate avatar experiences, especially romantic activities occurring through XR [13]. As early as 2008, in response to user frustrations with assessing compatibility with potential romantic partners through the standard dating website design, Frost and colleagues created a prototype VR environment that allowed users to go on "virtual dates" in a museum [4]. More recently in 2015 Zytko and Freeman envisioned different ways that augmented reality (AR) could scaffold face-to-face dates between users [10]. Now with the proliferation of mass market VR headsets and mobile devices these ideas for XR-enriched online dating do not have to stay relegated to prototypes and design concepts in the literature.

Near the end of 2021 Tinder announced plans for the "Tinderverse," or the leveraging of VR and AR technology for

"blurring the boundaries between offline and online" [8]. They are already beta testing a social VR environment for online daters in Seoul, Korea called Singletown in which users manipulate audio-enabled avatars and can go on virtual dates at a "piano bar" and other settings [2]. An in-app currency called Tinder Coins is also being beta tested in Europe [7] that the company claims will be used for purchasing "virtual goods." These announcements follow a succession of new interactive experiences within the dating app expedited by the COVID-19 pandemic such as Swipe Nights – an interactive in-app video series, and Explore – a new method of discovering nearby daters that is reminiscent of TikTok. Tinder is not the only dating app pursuing XR integration. Bumble has indicated plans to expand "in the metaverse and in the Web 3.0 world" [5], and new VR-only dating apps such as Planet Theta are emerging.

## 3 Challenges and Opportunities for Safety in the Dating Metaverse

Dating apps are well known as facilitators of sexual harm. They have been repeatedly linked to sexual violence such as rape [e.g., 9], as well as online sexual harms such as harassment [e.g, 1]. The inclusion of XR technologies into the online dating process may mitigate some of these harms. For example, prior work has reported on user struggles with assessing compatibility through dating apps [4], which has led some users to meeting face-to-face purely to collect more information about a potential dating partner [12]. Richer interaction afforded through social VR could allow users to bypass face-to-face encounters and the risks of harm that come with them. AR-equipped devices could also be used to help users navigate away from unsafe face-to-face encounters or direct bystanders to intervene in sexual harm occurring nearby (extending panic button-related functionality that we currently see in mobile apps).

XR-enabled dating apps may also exacerbate harm. For example, social VR environments like Tinder's Singletown will provide a new context for harm to occur. Prior HCI research has elucidated various types of harassment that users face in social VR environments including verbal harassment, virtual groping of their avatars, and environmental harms such as posting of lewd imagery [3,6]. These harms will surely translate—and likely escalate—in social VR environments intended for dating, laying a foundation for online harm more vivid and more severe than that that currently occurring over messaging and other relatively less-rich media used by online daters today.

Most concerning is the potential for sexual violence across virtual and physical worlds that XR-enabled dating apps may amplify. Our recent research at CSCW has demonstrated how experiences within dating apps can shape one's understanding of appropriate sexual behavior and sexual agency (their perceived right to decline a sexual advance) [11]. For example, the frequency of harassment in dating apps has lead users to perceive this behavior as normal, and the frequency of sexual requests in dating apps has developed a perceived "obligation" in some users to perform sexually when meeting other online daters face-to-face. Richer interaction afforded through XR can deepen the severity and frequency of such experiences, which can lead young and impressionable users to normalize the behavior and develop detrimental perceptions of sexual agency.

Our prior work has also elucidated ways in which users interpret content received in dating apps as signals of consent to sex – they subsequently make sexual advances during face-to-face encounters without confirming that their partner desires sex because they believe consent has already been conveyed online. Incorporation of social VR-based interaction prior to face-to-face meeting, as well as AR-enabled content during face-to-face interaction, can increase risk of users misinterpreting others' agreement to a sexual activity. Other concerns relating to safety are the virtual crimes relating by exposure of sensitive user data and inadequate privacy and security policies. [14]

Ultimately, given that users of current social VR environments as well as dating apps skew to younger age ranges, XR-enabled dating apps may inadvertently foster harmful practices and perceptions around sexual consent exchange if not designed to intentionally mitigate this influence.

## 4 Designing a Harm-Mitigative Dating Metaverse

Participatory design has long been championed within HCI research as a method to incorporate diverse stakeholders in technology design and decision-making. We are currently conducting a participatory design study with various stakeholders related to sexual violence to produce novel XR design patterns for preventing sexual harms in online dating. In line with best practices in public health, we pursue preventative solutions so as to help users avoid the significant psychological and physical trauma that is incurred through sexual violence.

Stakeholders that we are recruiting include sexual violence experts (nurse practitioners, operators of sexual violence shelters, and researchers of sexual violence) as well as demographics at disproportionate risk of sexual violence (e.g., women, members of the LGBTQIA+ community). Our participatory design sessions differ slightly based on the stakeholders. Sessions with sexual violence experts involve collaborative discussions to forecast sexual violence that may be facilitated through XR-enabled dating apps and expected antecedents/facilitators of such harm as informed by the expert's professional experience. Challenges of these sessions pertain to priming experts of XR technologies which they are typically unfamiliar with. Sessions with at-risk demographics involve reflections on their past experiences of sexual harm online, as well as storyboard scenarios to familiarize them with potential XR-enabled harms. Challenges with these sessions pertain to participant care: ensuring that the process of participation does not incur re-traumatization or otherwise cause harm to the participant.

After these initial activities we subsequently moderate design exercises for XR-enabled sexual violence mitigation by providing participants with various tools for creating physical constructions of potential XR design patterns, such as moldable clay, Lego sets, arts-and-crafts boxes, and "mechanic cards" (specially designed cards by our research team that each include a potential feature/idea that could be incorporated into the stakeholder's design). We have erred away from having participants design directly within VR environments due to reluctance of some participants—particularly sexual violence experts—to engage with technical tools for ideation.

## 5  Author Bios

**Sarath S. Shanker** is a PhD student in the Oakland HCI Lab, situated in the Department of Computer Science and Engineering at Oakland University. Having lived in eastern and western parts of the globe and experiencing various facets of both cultures, he is a proponent of Diversity and Inclusion in technology design. Both his research interests and personal advocacy work are tuned towards making technology-enabled social interactions safer, more equitable, and discrimination-free. His research typically uses participatory design to involve at-risk and marginalized demographics in the design of emerging technologies.

**Douglas Zytko** is an Assistant Professor in the Department of Computer Science and Engineering at Oakland University. He is also Director of the Oakland HCI Lab, a hub for interdisciplinary research into online-to-offline harm. The lab integrates researchers in human-computer interaction, XR, human-robot interaction, AI, psychology, and nursing to leverage emerging technologies for the prevention of harms that emerge through the combination of computer-mediated and face-to-face interaction. Most relevant to this workshop, Doug's research explores how XR technologies can be used to foster adoption of harm-mitigative behaviors, such as responsible drinking practices and consent exchange practices that mitigate sexual violence. His work typically leverages participatory design methods to put at-risk stakeholders in position to articulate new use cases for emerging technologies pursuant to safety. The research also coalesces harm experts in clinical and research contexts to produce theory-informed technologies for altering behavior of potential perpetrators.